\documentclass[aps,prb,amsmath,amssymb,twocolumn,showpacs]{revtex4-1}
\usepackage{bm}
\usepackage{graphicx}
\usepackage{epstopdf}
\usepackage{amsmath}
\usepackage[unicode=true,colorlinks=true]{hyperref}
\begin{document}
\title{Orbital magnetic ratchet effect}
\author{G. V. Budkin}
\author{L. E. Golub}
\affiliation{Ioffe Physical-Technical Institute of the Russian Academy of Sciences, 194021 St.~Petersburg, Russia}%

\begin{abstract}
Magnetic ratchets -- two-dimensional systems with superimposed non-centrosymmetric ferromagnetic gratings -- are considered theoretically. It is demonstrated that excitation by radiation results in a directed motion of two-dimensional carriers due to pure orbital effect of the periodic magnetic field. Magnetic ratchets based on various two-dimensional systems like topological insulators, graphene and semiconductor heterostructures are investigated. The mechanisms of the electric current generation caused by both radiation-induced heating of carriers and by acceleration in the radiation electric field in the presence of space-oscillating Lorentz force are studied in detail. The electric currents sensitive to the linear polarization plane orientation as well as to the radiation helicity are calculated. It is demonstrated that the frequency dependence of the magnetic ratchet currents is determined by the dominant elastic scattering mechanism of two-dimensional carriers and differs for the systems with linear and parabolic energy dispersions.
\end{abstract}

\maketitle

\section{Introduction}

Ratchets are periodic structures with broken spatial symmetry. 
A directed motion of particles is possible in these systems if they are driven out of thermal equilibrium even in the absence of an averaged force. Ratchets can be realized in various condense matter, biological and chemical systems which are intensively studied nowadays.~\cite{Hanggi_2009,Hanggi_2014} 
Advances in semiconductor nanotechnology allow fabricating various heterostructures with superimposed lateral superlattices which demonstrate ratchet effects due to lack of space inversion.~\cite{Olbrich_PRL_09,Olbrich_PRB_11,Review_JETP_Lett,Popov_Knap} The ratchets can be based on both traditional heterostructures and on graphene.~\cite{Nalitov,Kamann} 

The superlattice is often made by depositing metal grating above the two-dimensional structure, and recently semiconductor heterostructures with a grating of ferromagnetic stripes on top of the sample have been studied.~\cite{Weiss} These structures can be called \textit{magnetic ratchets}.
Ratchet effect has been demonstrated in the presence of a space-oscillating magnetic fields.~\cite{long_magn_ratchet,patterned_magn_film}
A ratchet effect purely magnetic in origin has been observed in a superconducting-magnetic nanostructure hybrid.~\cite{supercond}
Pure spin current is generated in systems with non-uniform magnetic fields.~\cite{Zeeman_ratchet} 
Possibility for large ratchet currents has been demonstrated in magnetic superlattices on the surface of topological insulators.~\cite{Lindner} However all studies on magnetic ratchets are focused on the Zeeman interaction of ferromagnetic metallic stripes with spins of two-dimensional carriers. 

\begin{figure}[t]
\includegraphics[width=0.9\linewidth]{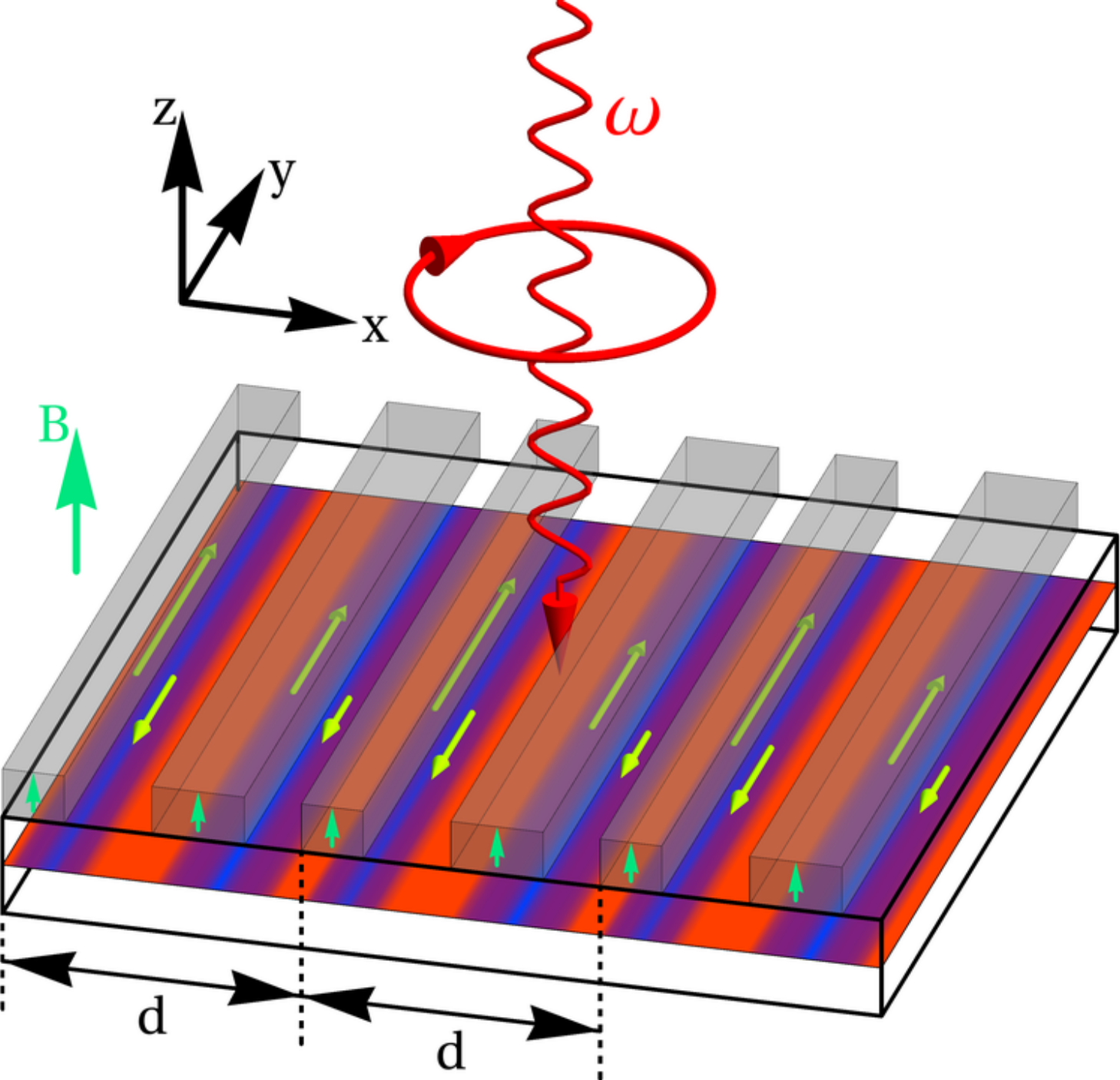}
\caption{Studied structure with normal space-oscillating magnetic field $\bm B(x)$
caused by non-centrosymmetric ferromagnetic grating with a period $d$.
At incidence of radiation of frequency $\omega$, the transmitted intensity is periodically modulated which results in space-oscillating carrier temperature shown by different colors.
If the magnetic field and the temperature gradient are in phase,  the ratchet current is excited due to the Nernst-Ettingshausen effect.}
\label{fig_structure}
\end{figure}

In this work we consider an orbital effect of space-periodic magnetic field. 
We investigate a two-dimensional structure with a superimposed non-centrosymmetric magnetic lateral superlattice. The magnetic field $\bm B$ is oriented normally to the two-dimensional plane and periodically changes with an in-plane coordinate $x$, Fig.~\ref{fig_structure}.  Under excitation of this system by electromagnetic radiation, the amplitude of the electric field acting on two-dimensional carriers, $E_0(x)$, is also periodically modulated in space. This is caused by, e.g., modulated reflection of the radiation from the metallic stripes. 
It is crucial for the ratchet effect that, due to lack of the space inversion the periodic functions $B(x)$ and $E_0^2(x)$ can have some phase difference.
The parameter controlling the existence of the magnetic ratchet current is given by
\begin{equation}
\label{asymm}
	\Xi=\overline{ B(x){d|E_0(x)|^2 \over dx} },
\end{equation}
where the bar denotes averaging over the $x$ coordinate.
If $\Xi$ is nonzero then the electric current is present. 

The studied system has a symmetry $C_s$ with only one reflection plane $(zx)$, and the normal component of the magnetic field changes its sign under reflection in this plane. The symmetry analysis yields the following expressions for the two-dimensional current density $\bm j$ generated in the magnetic ratchet under excitation by polarized radiation:
\begin{align}
\label{phenom}
&	j_x = \Xi \left [\chi_L (e_xe_y^*+e_ye_x^*) + \chi_C {\rm i} (e_xe_y^*-e_ye_x^*) \right] , \nonumber \\
&	j_y= \Xi \left [\chi_0 + \tilde{\chi}_L (|e_y|^2 - |e_x|^2) \right].
\end{align}
Here normally incident radiation with the electric field $\bm E = (E_0(x) \bm e \exp{(-{\rm i}\omega t)}+ c.c.)$ is considered with $\bm e$ being a complex polarization vector. We see that unpolarized radiation can generate the current perpendicular to the modulation direction ($j_y \propto \chi_0$) while under circular polarization the current along the $x$ axis ($j_x \propto \chi_C$) is excited. Linearly polarized radiation excites both $j_x \propto \chi_L$ and $j_y \propto \tilde{\chi}_L$ depending on the orientation of the polarization plane relative to the modulation direction.
Note that an average normal magnetic field and a homogeneous excitation do not result in the current generation in contrast to the in-plane field geometry.~\cite{Drexler_NNano}

In the next Section, we investigate the current caused by heating of the carriers.
In Sec.~\ref{Pol_depend}, we study the polarization-dependent currents which 
are caused by acceleration of carriers by the radiation electric field. Discussion of the results for magnetic ratchets based on semiconductor heterostructures, graphene and topological insulators
is given in Sec.~\ref{disc}, and Sec.~\ref{concl} concludes the paper.

\section{Nernst-Ettingshausen  ratchet}
\label{NE}

Under unpolarized excitation, the electric current is generated due to heating of the carriers. The absorbed intensity-modulated radiation creates a space-oscillating distribution of the electron temperature. In the presence of magnetic field, the electric current $j_y$ is generated locally due to the Nernst-Ettingshausen effect. 
 These currents flow in opposite directions in the areas with positive and negative temperature gradients, see yellow arrows in Fig.~\ref{fig_structure}.
If the modulated in the $x$ direction magnetic field is  in phase with the temperature gradient, so that the average~\eqref{asymm} is nonzero, then a net ratchet current is generated.
In this Section we derive the current caused by this mechanism which can be called the Nernst-Ettingshausen  ratchet current.

The radiation with the space-modulated intensity creates a periodic modulation of the electron temperature. 
The space-modulated correction to the temperature $\delta T(x)$ is found from the energy balance equation~\cite{Review_JETP_Lett,Nalitov}
\begin{equation}
\label{delta_T}
	{\delta T(x) \over \tau_\varepsilon} = 2|E_0(x)|^2 {e^2\tau_{\rm tr} v_{\rm F}/p_{\rm F} \over 1+(\omega\tau_{\rm tr})^2},
\end{equation}
where $\tau_\varepsilon$ is the energy relaxation time, $v_{\rm F}$, $p_{\rm F}$ are the Fermi velocity and Fermi momentum, and $\tau_{\rm tr}$ is the transport relaxation time.

The electric current density of the Nernst-Ettingshausen effect is given by~\cite{Anselm,Varlamov_Kavokin}
\begin{equation}
\label{j_NE}
		j_y^{\rm NE} = \left(\beta_{yx} - \beta_{xx} {\sigma_{yx} \over \sigma_{xx}} \right)  {d T \over dx},
\end{equation}
where $\hat{\bm \sigma}$ and $\hat{\bm \beta}$ are the conductivity and 
the thermoelectric
tensors, respectively.
For degenerate electrons with the Fermi energy $\varepsilon_{\rm F} \gg T$ we have
\begin{multline}
	\sigma_{xx} = {Ne^2 \tau_{\rm tr}v_{\rm F}\over p_{\rm F}},
	\quad 	
	\beta_{xx} = -{Ne^2 \pi^2 T\over 3 c p_{\rm F}^2} {d\left( \tau_{\rm tr}p_{\rm F}v_{\rm F}\right)\over d\varepsilon_{\rm F} } ,
\\
	\sigma_{yx} = - B {Ne^3 \tau_{\rm tr}^2 v_{\rm F}^2\over c p_{\rm F}^2} , \quad
\beta_{yx} = B {Ne^2 \pi^2 T\over 3 c p_{\rm F}^2} {d\left(\tau_{\rm tr}^2 v_{\rm F}^2\right)\over d\varepsilon_{\rm F} }.
\end{multline}
Here $N$ is the two-dimensional concentration.

Averaging the current density~\eqref{j_NE} over the structure period with account for Eq.~\eqref{delta_T} and the coordinate dependence of the magnetic field, we obtain the Nernst-Ettingshausen contribution to the magnetic ratchet current in the form 
${j_y= \Xi \chi_0^{\rm NE} }$ with $\Xi$ given by Eq.~\eqref{asymm} and 
\begin{equation}
\label{chi_0_NE}
	\chi_0^{\rm NE} = {2\pi^2 N e^4 T\tau_\varepsilon v_{\rm F}^2  \tau_{\rm tr}^2 \over 3 c p_{\rm F}^2 [1+(\omega\tau_{\rm tr})^2]} {d\over d\varepsilon_{\rm F}} \left({v_{\rm F}   \tau_{\rm tr} \over p_{\rm F}} \right)  .
\end{equation}
This equation demonstrates that 
$\chi_0^{\rm NE}$ has a Lorentzian frequency dependence.
However, the magnetic ratchet current is determined by energy dispersion and elastic scattering mechanism. 

For magnetic ratchets based on topological insulators or graphene, the carriers have linear energy dispersion $\varepsilon_p =v_0 p$, and the transport relaxation time is $\tau_{\rm tr}  \propto 1/\varepsilon_{\rm F}$ for scattering by short-range defects while for Coulomb impurity scattering $\tau_{\rm tr} \propto \varepsilon_{\rm F}$. It follows from Eq.~\eqref{chi_0_NE} that the Nernst-Ettingsgausen contribution is absent at Coulomb impurity scattering. In contrast, for short-range defects we have
\begin{equation}
\label{chi_0_NE_lin_disp}
		\chi_0^{\rm NE} = -{4\pi^2 N e^4 T\tau_\varepsilon v_0^2  \tau_{\rm tr}^3 \over 3 c p_{\rm F}^4 [1+(\omega\tau_{\rm tr})^2]} .
\end{equation}

The situation is opposite for parabolic energy dispersion realized in ratchets based on semiconductor heterostructures: The Nernst-Ettingsgausen contribution is zero at short-range scattering when $\tau_{\rm tr}$ is independent of the Fermi energy, while at Coulomb scattering with $\tau_{\rm tr} \propto \varepsilon_{\rm F}$ this contribution is given by Eq.~\eqref{chi_0_NE_lin_disp} where $v_0^2$ is substituted by $-v_{\rm F}^2$.

\section{Polarization-dependent magnetic ratchet}
\label{Pol_depend}

For calculation of the polarization-dependent currents we solve the Boltzmann kinetic equation for the distribution function $f_{\bm p}(x)$. 
We assume that the radiation photon energy $\hbar\omega$ is much smaller than $\varepsilon_{\rm F}$.
In order to account for the polarization of radiation we  consider its electric field as a force acting on the carriers together with the Lorentz force:
\begin{equation}
	\bm F_{\bm p}(x) = e [ \bm E(x) {\rm e}^{-{\rm i}\omega t} + c.c.] + {e\over c} \bm v_{\bm p} \times \bm B(x). 
\end{equation}
Here $\bm v_{\bm p}$ is the velocity of a carrier with the momentum $\bm p$.
The kinetic equation has the following form
\begin{equation}
\label{kin_eq}
	\left[ \partial_t + v_{\bm p,x}\partial_x + \bm F_{\bm p}(x) \cdot \bm \nabla_{\bm p} \right] f_{\bm p}(x) = St(f_{\bm p}),
\end{equation}
where the right-hand side is the elastic collision integral.

We assume that the electron mean-free path $v_{\rm F}\tau_{\rm tr}$ and the
energy-diffusion length $v_{\rm F}\sqrt{\tau_{\rm tr} \tau_\varepsilon}$ are small compared
with the superlattice period $d$. We also neglect the ac diffusion
provided $v_{\rm F} \ll \omega d$. On the other hand, no restrictions
are imposed on the value of $\omega\tau_{\rm tr}$. 
In the first order in the electric field the solution has the form
\begin{equation}
	f^{(1)}_{\bm p}(x) = {e (-df_0/dp)\tau_1(p)/p \over 1-{\rm i}\omega \tau_1(p)} \bm p \cdot \bm E(x)   + c.c.
\end{equation}
Here 
$f_0(p)$ is the Fermi-Dirac distribution function, 
and we introduce elastic relaxation times of the $n^{\rm th}$ angular harmonics of the distribution function
$\tau_n(p)$ 
(${n=1,2}$).
Note that the transport time ${\tau_{\rm tr}=\tau_1(p_{\rm F})}$.

We obtain the correction to the distribution function as a result of three more iterations of the kinetic equation accounting  the magnetic field $B(x)$, the space gradient $\partial_x$ and once more the radiation electric field $\bm E(x)$. The gradient should not be taken at the last stage because the result nullifies after averaging over the coordinates. Therefore, after substitution of $f^{(1)}_{\bm p}(x)$ into the kinetic Eq.~\eqref{kin_eq}, we obtain the four corrections which differ by the order of perturbations:
\[
\delta f^{(\partial_xEB)}_{\bm p}, \quad
\delta f^{(B\partial_xE)}_{\bm p}, \quad
\delta f^{(E\partial_xB)}_{\bm p}, \quad
\delta f^{(\partial_xBE)}_{\bm p}.
\]
The electric current density is given by
\begin{equation}
\label{j}
	\bm j = \nu e \sum_{\bm p} \bm v_{\bm p} \, \overline{\delta f_{\bm p}},
\end{equation}
where the factor $\nu$ accounts for the spin and valley degeneracy: $\nu =2,4$ and~1 for magnetic ratchets based on semiconductor heterostructures, graphene and topological insulators, respectively.
Correspondingly, we get four contributions to the ratchet current. 

Calculations show that all four contributions yield
\[
\tilde{\chi}_L=\chi_L,
\] 
i.e. the currents sensitive to the linear polarization plane orientation,  Eq.~\eqref{phenom}, have equal magnitudes.
The contributions to $\chi_L$, to the helicity-dependent current $\chi_C$ as well as to the polarization-independent current $\chi_0$ which is not related to heating are given by

\begin{subequations}
\label{qEB}
	\begin{multline}
		\chi_0^{(\partial_xEB)} = {N e^4 v_{\rm F} \tau_{\rm tr} \over 4 c  [1+(\omega\tau_{\rm tr})^2]} \\
		\times \left[ {1-\omega^2\tau_{\rm tr}\tau_2 \over 1+(\omega\tau_2)^2} 
		\tau_2  \left( {v_{\rm F}^2 \tau_{\rm tr}^2 \over p_{\rm F}^2}\right)' 
		- \tau_{\rm tr} {\left( v_{\rm F}^2 \tau_{\rm tr}^2 \right)' \over p_{\rm F}^2} \right],
	\end{multline}
	\begin{equation}
		\chi_L^{(\partial_xEB)} = {N e^4 v_{\rm F} \tau_{\rm tr}^2 \over 4 c p_{\rm F}^2 [1+(\omega\tau_{\rm tr})^2]}  \left( v_{\rm F}^2 \tau_{\rm tr}^2 \right)' ,
	\end{equation}
	\begin{multline}
		\chi_C^{(\partial_xEB)} = {N e^4 v_{\rm F} \tau_{\rm tr} \over 4 c  [1+(\omega\tau_{\rm tr})^2]} \\
		\times \left[ {\omega\tau_2(\tau_{\rm tr}+\tau_2) \over 1+(\omega\tau_2)^2} 
		 \left( {v_{\rm F}^2 \tau_{\rm tr}^2 \over p_{\rm F}^2}\right)' 
		-  {\left( v_{\rm F}^2 \tau_{\rm tr}^2 \right)' \over \omega p_{\rm F}^2}\right].
	\end{multline}
\end{subequations}
Hereafter the prime means differentiation over $p_{\rm F}$.

The second contribution is given by
\begin{subequations}
\label{BqE}
\begin{equation}
	\chi_L^{(B\partial_xE)} = {N e^4 v_{\rm F}^2 \tau_{\rm tr}^3 \left( p_{\rm F} v_{\rm F} \tau_{\rm tr}\right)' \over 2 c p_{\rm F}^3 [1+(\omega\tau_{\rm tr})^2]^2},
\end{equation}
\begin{multline}
	\chi_0^{(B\partial_xE)} = \chi_L^{(B\partial_xE)} +  
	{N e^4 v_{\rm F}^2 \tau_{\rm tr}^2 \over 4 c p_{\rm F} [1+(\omega\tau_{\rm tr})^2]^2} \\
\times	{1-\omega^2\tau_{\rm tr}(\tau_{\rm tr}+2\tau_2) \over 1+(\omega\tau_{\rm tr})^2} \tau_2  \left( {v_{\rm F} \tau_{\rm tr} \over p_{\rm F} }\right)',
\end{multline}
\begin{multline}
	\chi_C^{(B\partial_xE)} = {N e^4 v_{\rm F}^2 \tau_{\rm tr}^2 \over 4 c p_{\rm F} [1+(\omega\tau_{\rm tr})^2]^2} \\
	\times \Biggl[ {\omega\tau_2 (2\tau_{\rm tr}+\tau_2 - \omega^2\tau_{\rm tr}^2\tau_2) \over 1+(\omega\tau_2)^2}  \left( {v_{\rm F} \tau_{\rm tr} \over p_{\rm F} }\right)' \\
	- {(1-\omega^2\tau_{\rm tr}^2) \left( p_{\rm F} v_{\rm F} \tau_{\rm tr}\right)' \over \omega p_{\rm F}^2}
	\Biggr].
\end{multline}
\end{subequations}

The next contribution is sensitive only to the linear polarization:
\begin{equation}
\label{EqB}
	\chi_L^{(E\partial_xB)} = - {N e^4  \tau_{\rm tr} \over 2 c p_{\rm F}^3 [1+(\omega\tau_{\rm tr})^2]} 
	\left( p_{\rm F} v_{\rm F}^3 \tau_{\rm tr}^2\tau_2\right)',
\end{equation}
while $\chi_0^{(E\partial_xB)} =\chi_C^{(E\partial_xB)}=0$.
In contrast, the fourth contribution does not change at rotation of the linear polarization plane ($\chi_L^{(\partial_xBE)}=0$) but has a polarization-independent contribution and one sensitive to the radiation helicity:
\begin{subequations}
\label{qBE}
\begin{equation}
	\chi_0^{(\partial_xBE)} = {N e^4 v_{\rm F}^2 \tau_{\rm tr} \tau_2^2 [1-\omega^2\tau_2(2\tau_{\rm tr}+\tau_2)] \over 2 c p_{\rm F} [1+(\omega\tau_{\rm tr})^2][1+(\omega\tau_2)^2]^2} 
	\left( {v_{\rm F} \tau_{\rm tr} \over p_{\rm F} }\right)',
\end{equation}
\begin{equation}
	\chi_C^{(\partial_xBE)} = {N e^4 v_{\rm F}^2 \tau_{\rm tr} \tau_2^2 \omega(\tau_{\rm tr}+2\tau_2-\omega^2\tau_{\rm tr}\tau_2^2) \over 2 c p_{\rm F} [1+(\omega\tau_{\rm tr})^2][1+(\omega\tau_2)^2]^2} 
	\left( {v_{\rm F} \tau_{\rm tr} \over p_{\rm F} }\right)'.
\end{equation}
\end{subequations}
The relaxation time $\tau_2$ is present because, at the intermediate stages of iterations of the kinetic equation, we obtained not only the first but also the second angular harmonics of the distribution function.~\cite{PDE_CR_Budkin}

The above derived expressions are valid for any energy dispersion of the two-dimensional carriers and for arbitrary dependence of the scattering times on the Fermi wavevector.
In the next Section we analyze the obtained results for magnetic ratchets based on two-dimensional systems with a linear energy dispersion like topological insulators or graphene and on semiconductor heterostructures with a parabolic dispersion.

\section{Discussion}
\label{disc}

The results of the previous Section demonstrate an existence of a ratchet current independent of the radiation polarization state, $\chi_0$, which is not related to the heating of carriers. Comparing these results with Eqs.~\eqref{chi_0_NE},~\eqref{chi_0_NE_lin_disp} we obtain an estimate for the ratio of the Nernst-Ettingshausen and elastic-scattering contributions to $\chi_0$:
\[ 
{\chi_0^{\rm NE} \over \chi_0} \sim \pi^2 {T\over \varepsilon_{\rm F}} {\tau_\varepsilon \over \tau_{\rm tr}}.
\]
At liquid Helium  temperature, the energy relaxation time of two-dimensional carriers has an order of nanoseconds, while the transport scattering time is $\tau_{\rm tr} \sim 1$~ps. Therefore, the Nernst-Ettingshausen contribution to
the polarization-independent current dominates at low temperatures for scattering by short-range defects in ratchets based on topological insulators or graphene and for scattering by smooth Coulomb potential in semiconductor heterostructures. However, in two opposite cases $\chi_0^{\rm NE}=0$, see Sec.~\ref{NE}, so the elastic-scattering contribution is dominant.

Equations derived for $\chi_i$ ($i=0,L,C$) which are the sums of four contributions
\begin{equation}
	\chi_i = \chi_i^{(\partial_xEB)}+\chi_i^{(B\partial_xE)}+\chi_i^{(E\partial_xB)}+\chi_i^{(\partial_xBE)}
\end{equation}
demonstrate that the frequency dependencies of the magnetic ratchet currents strongly depend on both the carrier energy dispersion and dominant elastic scattering mechanism.

\begin{figure}[t]
\includegraphics[width=0.9\linewidth]{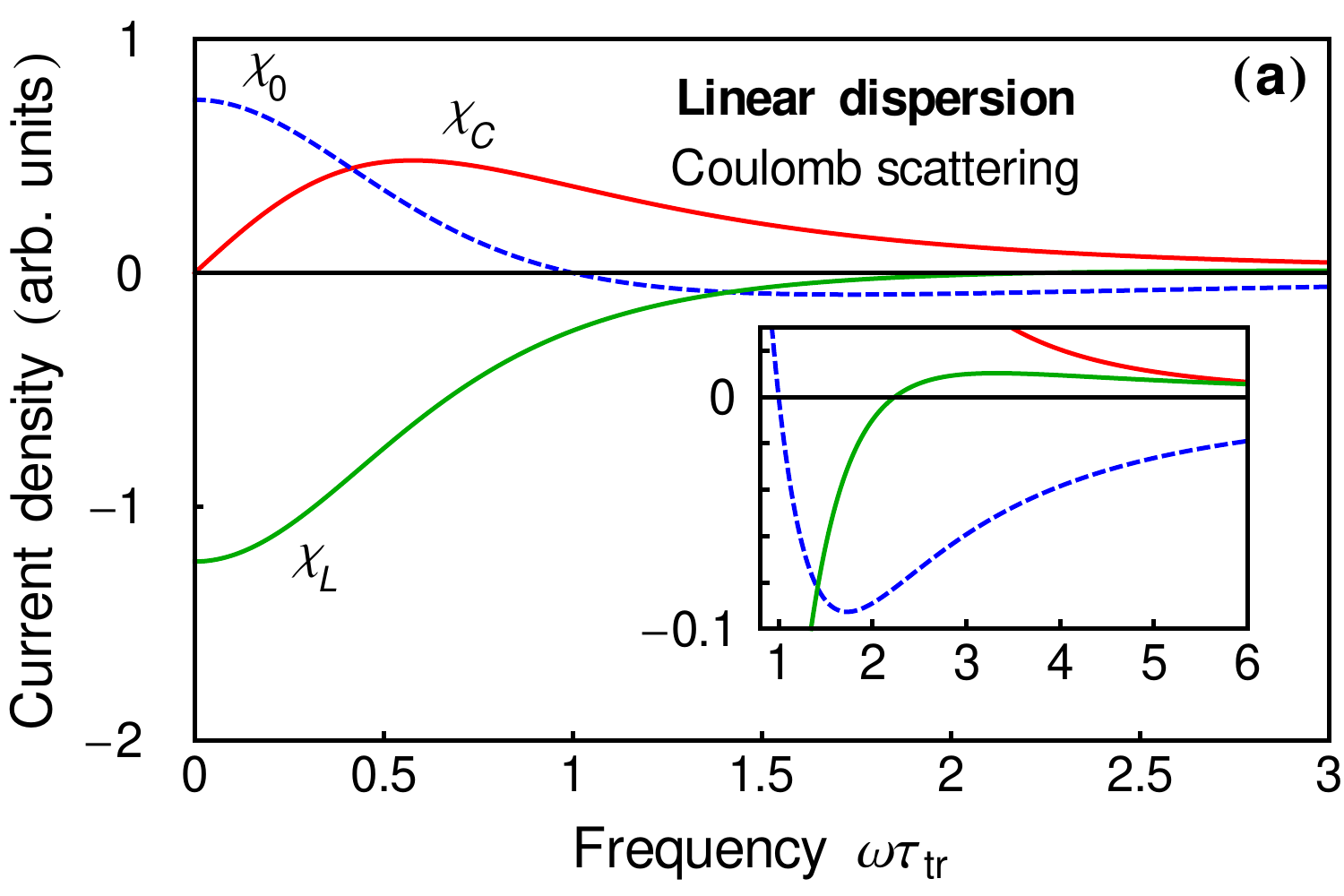}
\includegraphics[width=0.9\linewidth]{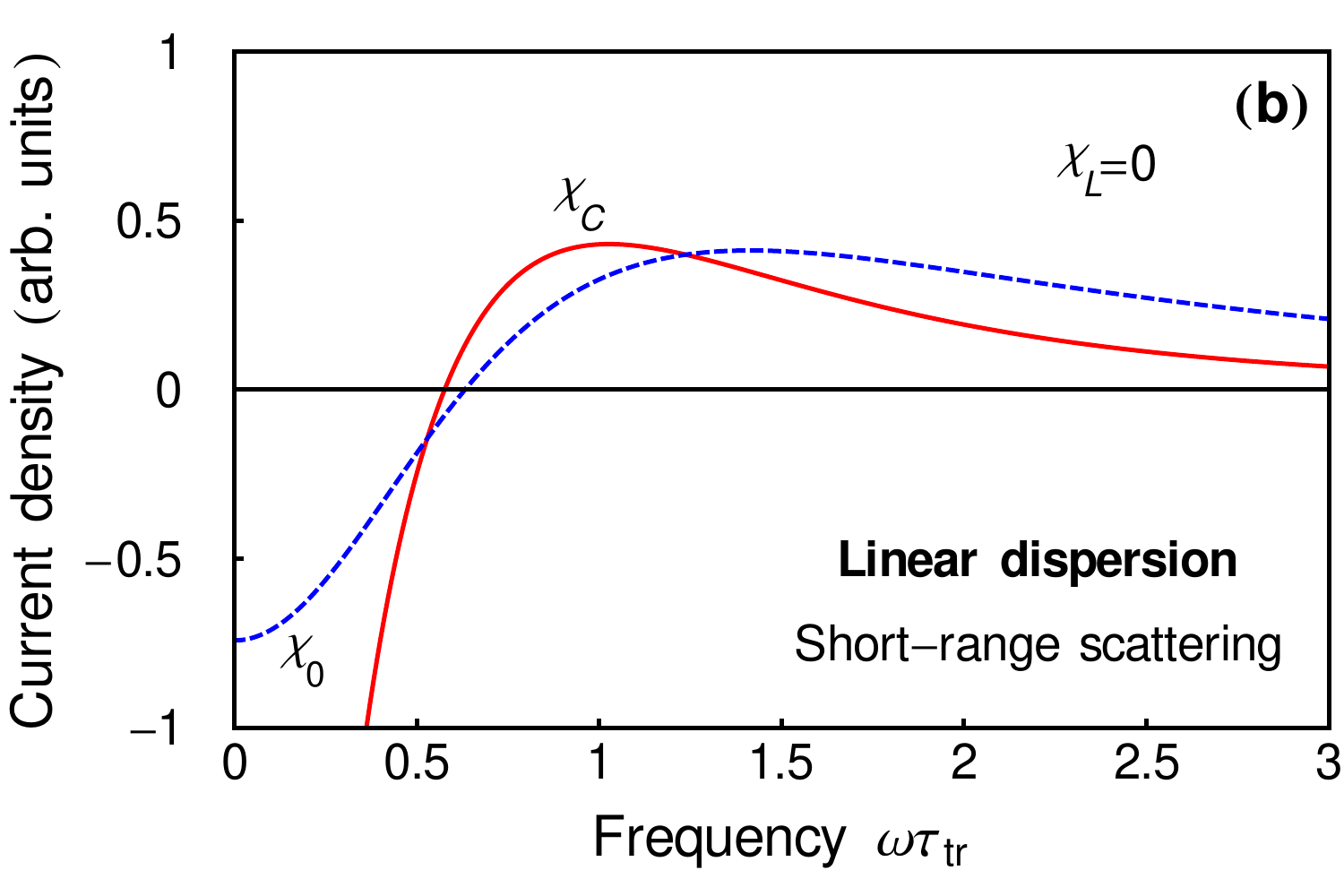}
\caption{Magnetic ratchet currents in the systems based on topological insulators or graphene at Coulomb impurity scattering (a) and at scattering by short-range impurities (b). Inset highlights the region where $\chi_L$ changes its sign.}
\label{fig_lin_disp}
\end{figure}

For magnetic ratchets based on topological insulators or graphene, the energy dispersion of two-dimensional carriers is linear.
We consider two most actual elastic scattering mechanisms:
scattering by Coulomb impurities and by short-range defects. In the case of Coulomb impurities, the elastic scattering times have the following dependence on the Fermi momentum:
\[
\tau_{\rm tr} = 3\tau_2 \propto p_{\rm F}.
\]
Figure~\ref{fig_lin_disp} shows frequency dependencies of the magnetic ratchet currents for a linear energy dispersion.
At Coulomb impurity scattering, the polarization-independent current $\propto \chi_0$ reverses its direction at ${\omega \approx \tau_{\rm tr}^{-1} }$. The current sensitive to the linear polarization degree $\propto \chi_L$ nullifies as well but it occurs at $\omega \tau_{\rm tr} \approx 2$. In contrast, the radiation-helicity sensitive contribution $\chi_C$ is sign-constant but it has a maximum at $\omega \tau_{\rm tr} \approx 0.6$, Fig.~\ref{fig_lin_disp}a. 

At scattering by short-range defects the situation differs drastically. 
The relaxation times have the following dependence on the Fermi momentum:
\[
\tau_{\rm tr} = 2\tau_2 \propto 1/p_{\rm F}.
\]
As a result, the linear-polarization sensitive ratchet current is absent: It follows from Eqs.~\eqref{qEB}-\eqref{EqB} that the contribution ${\chi_L^{(B\partial_xE)} \propto (p_{\rm F}\tau_{\rm tr})' =0}$, and two other contributions, $\chi_L^{(E\partial_xB)}$ and $\chi_L^{(E\partial_xB)}$, exactly cancel each other.  The polarization-independent contribution 
changes its sign at $\omega \tau_{\rm tr} \approx 0.6$ and has a maximum at $\omega \tau_{\rm tr} \approx 1.5$, Fig.~\ref{fig_lin_disp}b. The helicity-sensitive contribution $\chi_C$ has a maximum at $\omega \tau_{\rm tr} \approx 1$ in this case. At  $\omega \to 0$ it tends to zero since the circular-polarization dependent effects can not be present in the static limit. This occurs at $\omega \sim \tau_\varepsilon^{-1} \ll \tau_{\rm tr}^{-1}$, see Ref.~\onlinecite{Nalitov} for details.

\begin{figure}[t]
\includegraphics[width=0.9\linewidth]{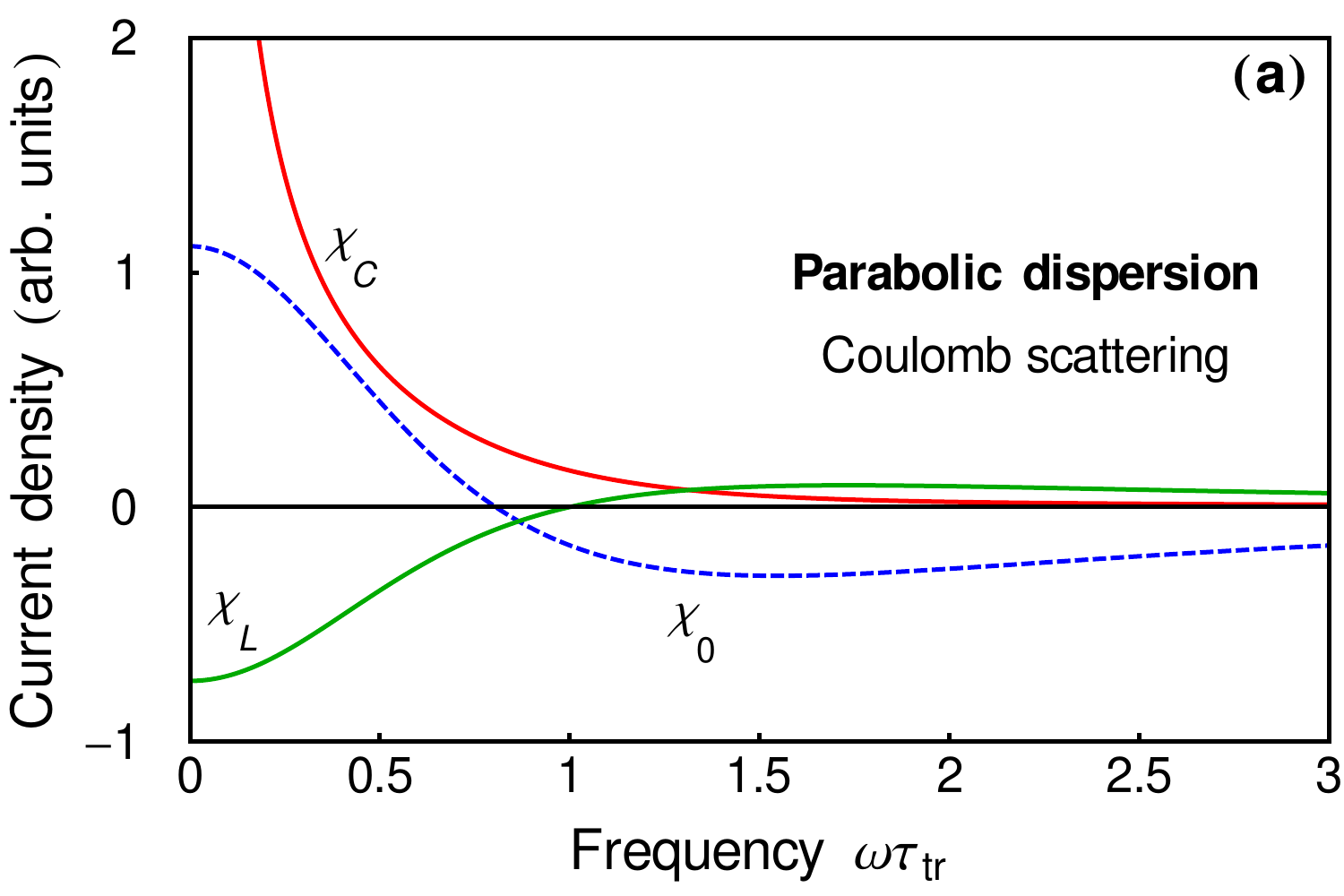}
\includegraphics[width=0.9\linewidth]{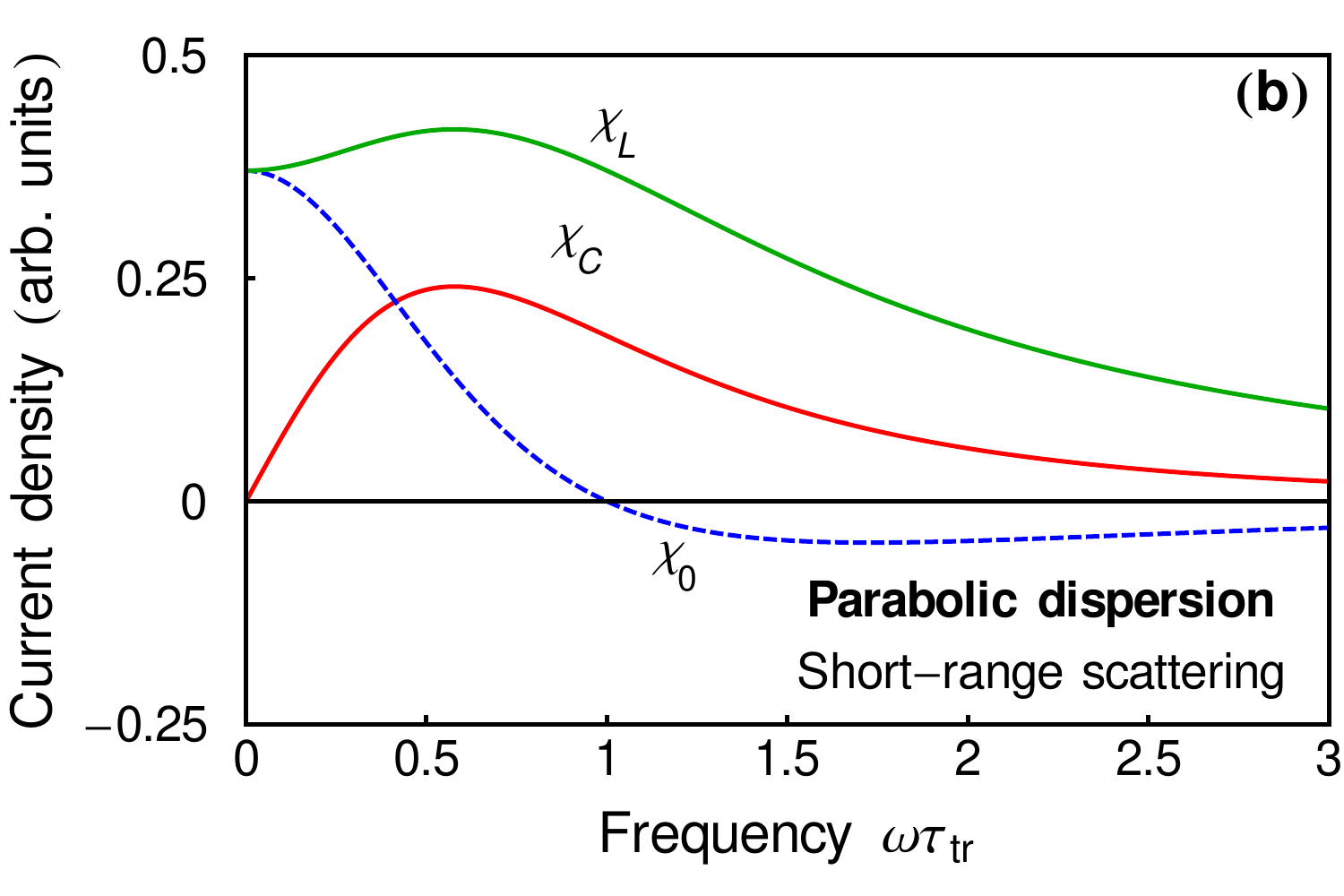}
\caption{Current in magnetic ratchets based on heterostructures with a parabolic energy dispersion at Coulomb impurity scattering~(a) and scattering by short-range defects~(b).}
\label{fig_parab_disp}
\end{figure}

Now we turn to magnetic ratchets based on semiconductor heterostructures. In this case the energy dispersion is parabolic, and for Coulomb impurity scattering we have:
\[
\tau_{\rm tr} = 2\tau_2 \propto p_{\rm F}^2.
\]
Substituting this into Eqs.~\eqref{qEB}-\eqref{qBE} we obtain nonzero results for all ratchet currents. The frequency dependencies are shown in Fig.~\ref{fig_parab_disp}a. Both $\chi_0$ and $\chi_L$ change their signs at $\omega \approx \tau_{\rm tr}^{-1}$.
The helicity-dependent ratchet current $\propto \chi_C$ behaves as $1/\omega$ at both large and small $\omega\tau_{\rm tr}$ and tends to zero at low frequencies $\omega \sim \tau_\varepsilon^{-1}$.

Finally, at scattering by short-range defects we have for two-dimensional carriers with a parabolic energy dispersion:
\[
\tau_{\rm tr} = \tau_2,
\]
and both scattering times are independent of $p_{\rm F}$. In this case, the ratchet currents have approximately twice smaller amplitude than at Coulomb impurity scattering, Fig.~\ref{fig_parab_disp}b. The polarization-independent contribution $\chi_0$ has a similar frequency behaviour while $\chi_L$ increases in the frequency range $0 < \omega < \tau_{\rm tr}^{-1}$ and then decreases as $1/\omega^2$ at high frequencies. The helicity-dependent contribution $\chi_C$ has a maximum at $\omega\tau_{\rm tr} \approx 0.6$ and then drops as $1/\omega$.

In this work we assume that characteristic frequency range is far from plasmon frequency in the two-dimensional gas.
In resonant situation the ratchet current in the similar structures with non-magnetic grating is shown to be enhanced.~\cite{Popov_Knap} We also expect some new features in the magnetic ratchets where magnetoplasmons can be effectively excited.

\section{Conclusion}
\label{concl}

In conclusion,  we have demonstrated a possibility of the orbital magnetic ratchet effect.
The microscopic theory of this phenomenon is developed for the structures based on topological insulators, graphene and semiconductor heterostructures.
The Nernst-Ettingsgausen ratchet current is shown to exist only at short-range scattering in the systems with a linear energy dispersion and at Coulomb impurity scattering for parabolic dispersion. In the opposite cases the polarization-independent current is not related to the heating of carriers. The ratchet currents sensitive to the linear and circular polarization of radiation exist at any elastic scattering potential but their frequency dependencies 
differ strongly for
systems with linear and parabolic dispersions.

\acknowledgments 
We thank E.L. Ivchenko for helpful discussions. 
The work is supported by RFBR and the ``Dynasty'' Foundation.


\begin{thebibliography}{99}

\bibitem{Hanggi_2009}  P. H\"anggi and F. Marchesoni, Rev. Mod. Phys. \textbf{81}, 387 (2009).

\bibitem{Hanggi_2014}
S. Denisov, S. Flach, and P. H\"anggi, Phys. Rep. \textbf{538}, 77 (2014).

\bibitem{Olbrich_PRL_09} P. Olbrich, E. L. Ivchenko, R. Ravash, T. Feil, S. D. Danilov, J. Allerdings, D. Weiss, D. Schuh, W. Wegscheider, and S. D. Ganichev, Phys. Rev. Lett. \textbf{103}, 090603 (2009).

\bibitem{Olbrich_PRB_11} P. Olbrich, J. Karch, E. L. Ivchenko, J. Kamann, B. M\"{a}rz, M. Fehrenbacher, D. Weiss, and S. D. Ganichev, Phys. Rev. B \textbf{83}, 165320 (2011).

\bibitem{Review_JETP_Lett} E.L. Ivchenko and S. D. Ganichev, Pisma v ZhETF \textbf{93}, 752 (2011) [JETP Lett. \textbf{93}, 673 (2011)].

\bibitem{Popov_Knap}
V. V. Popov, D. V. Fateev, T. Otsuji, Y. M. Meziani, D. Coquillat,
and W. Knap, Appl. Phys. Lett. \textbf{99}, 243504 (2011).

\bibitem{Nalitov} 
A. V. Nalitov, L. E. Golub, and E. L. Ivchenko, Phys. Rev. B \textbf{86}, 115301 (2012).

\bibitem{Kamann} 
J. Kamann, J. Munzert, L.E. Golub,  M. K\"onig, J. Eroms, M. Mittendorff, S. Winnerl, F. Fromm, Th. Seyller, D. Weiss, and S.D. Ganichev, Proc. 32nd Int. Conf. Phys. Semicond. ICPS-32 (August 10-15, 2014, Austin, Texas, USA), in press.

\bibitem{Weiss} 
C. Betthausen, T. Dollinger, H. Saarikoski, V. Kolkovsky, G. Karczewski, T. Wojtowicz, K. Richter, D. Weiss, 
Science \textbf{337}, 324 (2012).

\bibitem{long_magn_ratchet} 
P. Tierno, P. Reimann, T. H. Johansen, and F. Sagu\'{e}s,
Phys. Rev. Lett. \textbf{105}, 230602 (2010).

\bibitem{patterned_magn_film} 
A. V. Straube and P. Tierno, EPL \textbf{103}, 28001 (2013).

\bibitem{supercond} 
D. Perez de Lara, F. J. Casta\~no, B. G. Ng, H. S. Korner, R. K. Dumas, E. M. Gonzalez, K. Liu, C. A. Ross, I. K. Schuller, and J. L. Vicent, Phys. Rev. B \textbf{80}, 224510 (2009).

\bibitem{Zeeman_ratchet} 
M. Scheid, D. Bercioux and K. Richter,
New J. of Phys. \textbf{9}, 401 (2007).

\bibitem{Lindner} 
N.H. Lindner, G. Refael, and F. von Oppen, arXiv:1403.0010.

\bibitem{Drexler_NNano} 
C. Drexler,	 S. A. Tarasenko,	 P. Olbrich,	 J. Karch,	 M. Hirmer,	 F. Müller,	 M. Gmitra,	 J. Fabian,	 R. Yakimova,	 S. Lara-Avila,	 S. Kubatkin,	 M. Wang,	 R. Vajtai,	 P. M. Ajayan,	 J. Kono,	 and S. D. Ganichev,
Nat. Nanotech. \textbf{8}, 104 (2013).


\bibitem{Anselm} A. I. Anselm, Introduction to Semiconductor Theory, 2nd edition (Prentice-Hall, Englewood Cliffs, NJ) 1982.
\bibitem{Varlamov_Kavokin} A. A. Varlamov and A. V. Kavokin, EPL \textbf{103}, 47005 (2013).


\bibitem{PDE_CR_Budkin} 
S. Stachel, G. V. Budkin, U. Hagner, V. V. Bel'kov, M. M. Glazov, S. A. Tarasenko, S. K. Clowes, T. Ashley, A. M. Gilbertson, and S. D. Ganichev, Phys. Rev. B \textbf{89}, 115435 (2014).


\end{thebibliography}
\end{document}